\documentclass[11pt]{article}
\usepackage{moriond,epsfig}

\def\be{\begin{equation}}
\def\ee{\end{equation}}
\def\bea{\begin{eqnarray}}
\def\eea{\end{eqnarray}}

\begin{document}
\vspace*{4cm}
\title{Estimation of SM backgrounds to SUSY searches in the 1-lepton + jets + \boldmath{$E_{\rm{T}}^{\rm{miss}}$} channel}

\author{ J. M. Lorenz \\
on behalf of the ATLAS Collaboration}

\address{Fakult\"at f\"ur Physik, Ludwig-Maximilians-Universit\"at-M\"unchen\\
Am Coulombwall 1, D-85748 Garching, Germany}

\maketitle\abstracts{
The ATLAS Collaboration has reported the first results of the search for SUSY particles in 1-lepton + $\geq$ 3 jets + $E_T^{\rm{miss}}$ final states for $\int L dt = 35 ~\rm{pb^{-1}}$.
An essential ingredient for these results is a reliable background estimation in the signal region, in particular of the $t\bar{t}$, \textit{W}+jets and QCD backgrounds. The estimation of these three backgrounds is explained in this paper. The $t\bar{t}$ and \textit{W}+jets backgrounds are obtained from a background dominated control region and extrapolated to the signal region, whereas for the estimation of the QCD background a matrix method is used.}

\section{Introduction}

\noindent Supersymmetric extensions (SUSY) to the Standard Model (SM) predict the existence of supersymmetric particles which could be produced in the proton-proton collisions at a center-of-mass energy of 7 TeV at the LHC \cite{LHCcol}. The search for SUSY is one of the main aims of the ATLAS experiment \cite{ATLASexp}. 

\noindent The main production channels of SUSY particles at the LHC are squark-(anti)squark, gluino-squark and gluino-gluino pairs if these sparticles are light enough. Typical squark and gluino decays contain isolated leptons, quarks and gluons (which result in jets) and end with the stable Lightest Supersymmetric Particle (LSP) if a R-parity conserving SUSY model is assumed. As the LSP escapes the detector undetected, missing transverse energy ($E_{\rm{T}}^{\rm{miss}}$) will be observed. Therefore, the typical experimental signature consists of multiple jets, isolated leptons and $E_{\rm{T}}^{\rm{miss}}$. Here only final states with one isolated muon or electron, at least three jets and $E_{\rm{T}}^{\rm{miss}}$ are considered (1-lepton + $\geq$ 3jets + $E_{\rm{T}}^{\rm{miss}}$ channel). 

\noindent However, other physical processes can have similar experimental signatures, in particular:
\begin{itemize}
 \item $t\bar{t}$ events with a semileptonic decay topology, where each top quark decays into a \textit{W}-boson and a b-quark, with one of
the \textit{W}-bosons decaying into an isolated charged lepton and a neutrino ($E_{\rm{T}}^{\rm{miss}}$).
 \item \textit{W}+jets events where the \textit{W}-boson decays into a neutrino and the corresponding charged lepton.
 \item QCD events, like heavy flavor (b or c quarks) decays, events with photon conversions or jets which were reconstructed as isolated leptons.
\end{itemize}

\begin{figure}
\begin{center}
\includegraphics[width=0.45\textwidth]{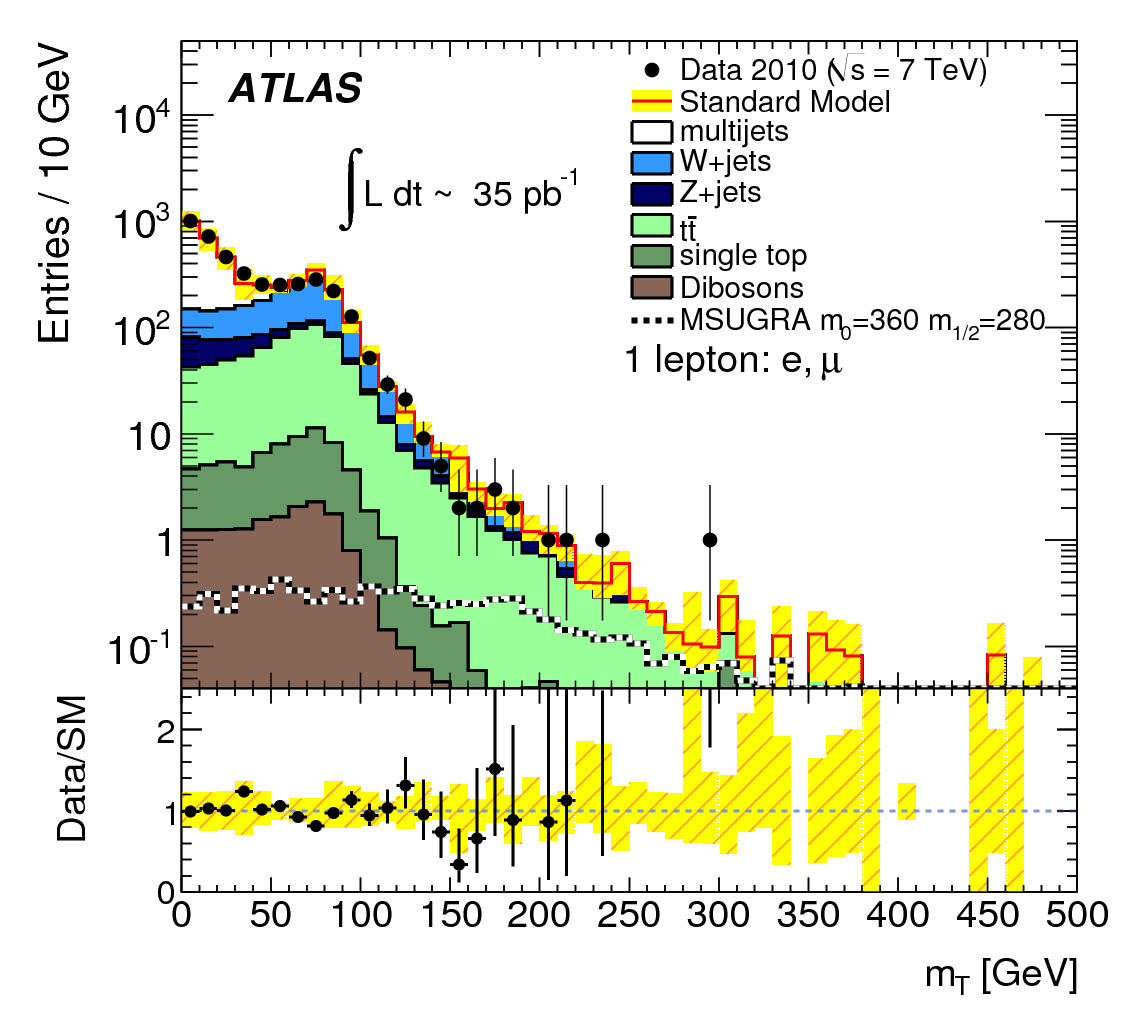}
\end{center}
\caption{\label{fig:mtcut} The $m_{\rm{T}}$ distribution after the lepton and the jets criteria. The ratio of data to SM expectations is given in the ratio plot below. The uncertainties on the MC predictions are indicated by the yellow bands.}
\end{figure}

\noindent A signal region rich in SUSY signal events but poor in background events (BG) which was motivated by studies on Monte Carlo event samples is defined as follows. Hereby, the channel itself is defined by asking for exactly one isolated lepton being in the event with a transverse momentum of $p_{\rm{T}} > 20~\rm{GeV}$ which already reduces the QCD BG.
Furthermore, the event must contain at least three jets with $p_{\rm{T}}^{\rm{leading~jet}} > 60~\rm{GeV}$ and $p_{\rm{T}}^{\rm{second,third~jet}} > 30~\rm{GeV}$. 
In addition to these cuts, various cuts are applied to reject background events: a further reduction of the QCD background is achieved by asking for the three leading jets and $E_{\rm{T}}^{\rm{miss}}$ not to point in the same direction \footnote{$\Delta \phi (\rm{jet}_i, \vec{E}_{\rm{T}}^{miss}) > 0.2 ~(i =1,2,3)$}. Furthermore, only events with $E_{\rm{T}}^{\rm{miss}} > \max(125~\rm{GeV}, 0.25 M_{\rm{eff}})$ \footnote{the effective mass is defined as $M_{\rm{eff}} = p_{\rm{T}}^{l} + E_{\rm{T}}^{\rm{miss}} + \sum_{i=1}^{3} p_T^{\rm{jet}_i}$} and a high transverse mass with $m_{\rm{T}} > 100~\rm{GeV}$ \footnote{the transverse mass is defined as $m_{\rm{T}} = \sqrt{2 \cdot p_{\rm{T}}^{l} \cdot E_{\rm{T}}^{\rm{miss}} \cdot (1 -\cos{(\Delta \phi(l,E_{\rm{T}}^{\rm{miss}}))})}$} are selected. The last cut reduces the \textit{W}+jets and $t\bar{t}$ backgrounds considerably as illustrated in figure \ref{fig:mtcut}. Finally, a cut on the effective mass with $M_{\rm{eff}} > 500~\rm{GeV}$ is applied.
After applying these cuts, $t\bar{t}$ events are the main background and QCD events are heavily suppressed.

\noindent The estimation of the three most important backgrounds in the signal region is discussed in the following.

\section{\textit{W} and top backgrounds}

\begin{figure}
\begin{center}
   \includegraphics[width=0.5\textwidth]{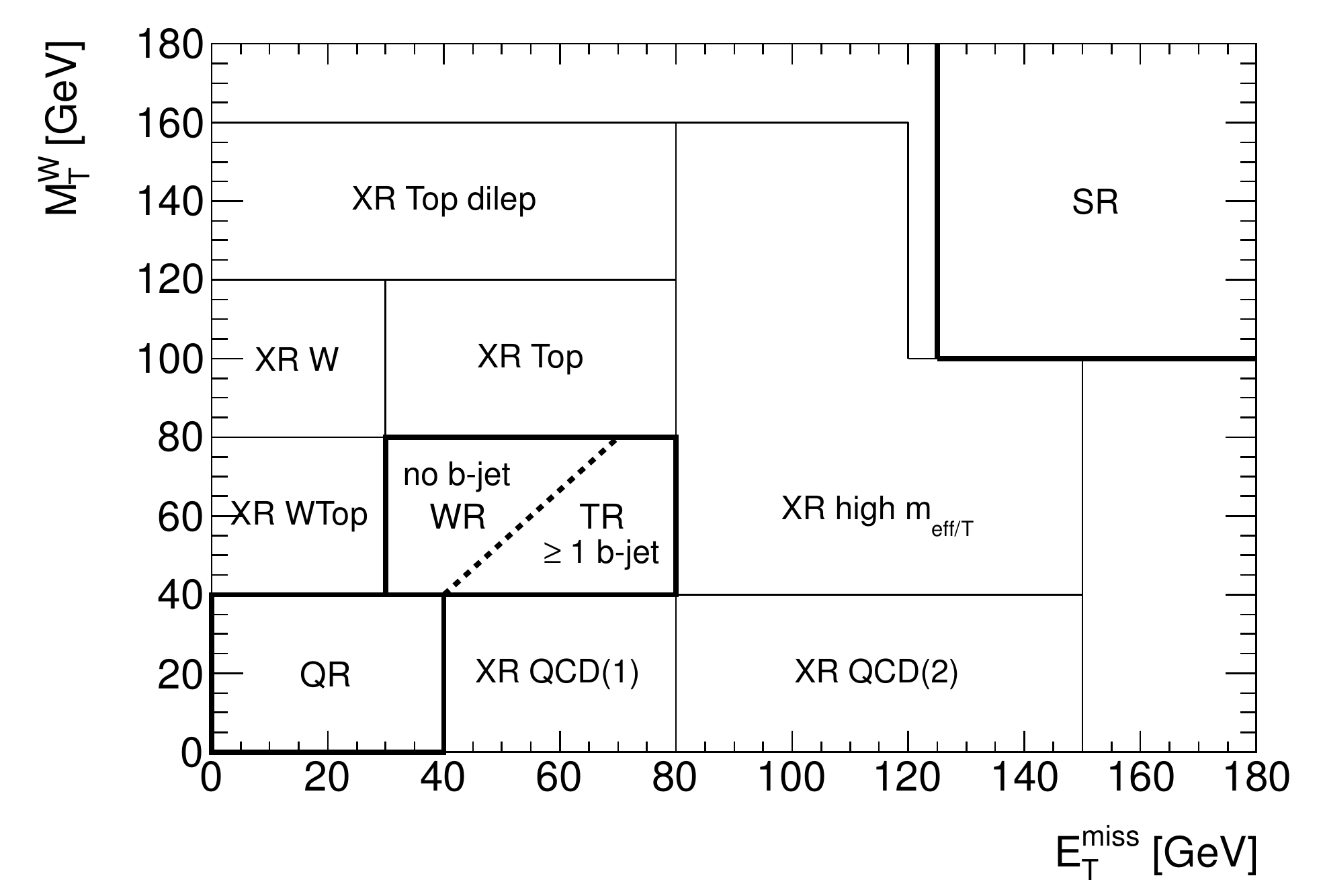}
\end{center}
\caption{\label{fig:control} The positions of the signal region (SR) and of the main control regions (as QR, WR and TR, defined in the text) are indicated in the $E_{\rm{T}}^{\rm{miss}}$ - $m_{\rm{T}}$-plane. Control regions which are not mentioned in the text (XR) are used to cross check the background estimation techniques presented in the text.}
\end{figure}

\noindent The magnitude of the backgrounds is estimated with the help of background dominated control regions. In figure \ref{fig:control}, various control regions - one for each of the main backgrounds (QCD, $t\bar{t}$, \textit{W}+jets) in the $E_{\rm{T}}^{\rm{miss}}$ - $m_{\rm{T}}$ plane are defined. All other cuts for these control regions correspond to the usual selection cuts presented above except the cut on $M_{\rm{eff}}$ which is not applied. For example, the QCD region (QR) which is defined by $E_{\rm{T}}^{\rm{miss}} < 40~\rm{GeV}$ and $m_{\rm{T}} < 40~\rm{GeV}$ is rich in QCD events. The control region with medium $E_{\rm{T}}^{\rm{miss}}$ and $m_{\rm{T}}$ values ($30 < E_{\rm{T}}^{\rm{miss}} < 80~\rm{GeV}$ and $40 < m_{\rm{T}} < 80~\rm{GeV}$) is dominated by top and \textit{W}+jets events. This control region is further divided into two control regions dominated by $t\bar{t}$ events (top region, TR) or \textit{W}+jets events (\textit{W} region, WR), respectively, by the requirement of finding at least one b-tagged jet (TR) or no b-tagged jet (WR) in the three leading jets. The QR is only used to estimate the QCD contamination in the TR or in the WR, whereas the QCD BG in the signal region itself is obtained with the method described in the next section. In contrast, the $t\bar{t}$ and \textit{W}+jets BG in the signal region is obtained by extrapolating the number of measured \textit{W} and top events in W and T control regions (other backgrounds were subtracted) into the signal region. For this an extrapolation factor which is obtained from Monte Carlo is used. The extrapolation is illustrated in equation \ref{eq:extrapolation} for the top background. 

\begin{scriptsize}
\be
\label{eq:extrapolation}
 \underbrace{N(t\bar{t}~\rm{pred.,}~ \rm{~SR})}_{\textnormal{predicted events in signal region}} = \underbrace{(N(t\bar{t} \rm{(data),~CR}))}_{\textnormal{measured events in control region - other BG}} \times \underbrace{\frac{N(t\bar{t}\rm{(MC),~SR})}{N(t\bar{t}\rm{(MC),~CR})}}_{\textnormal{extrapolation factor CR to SR}} \nonumber
\ee
\end{scriptsize}

\begin{figure}
\begin{center}
 \includegraphics[width=0.45\textwidth]{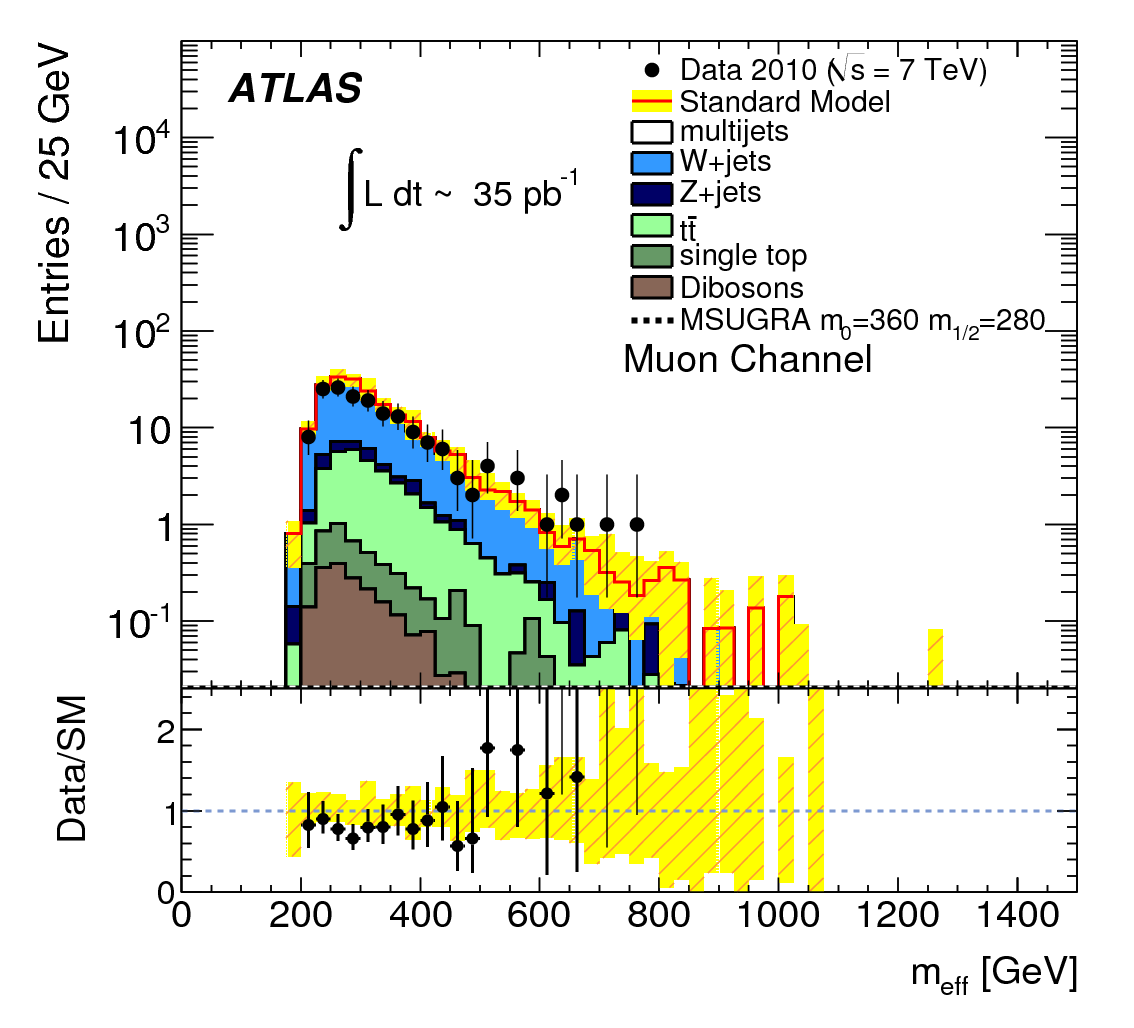} 
 \includegraphics[width=0.45\textwidth]{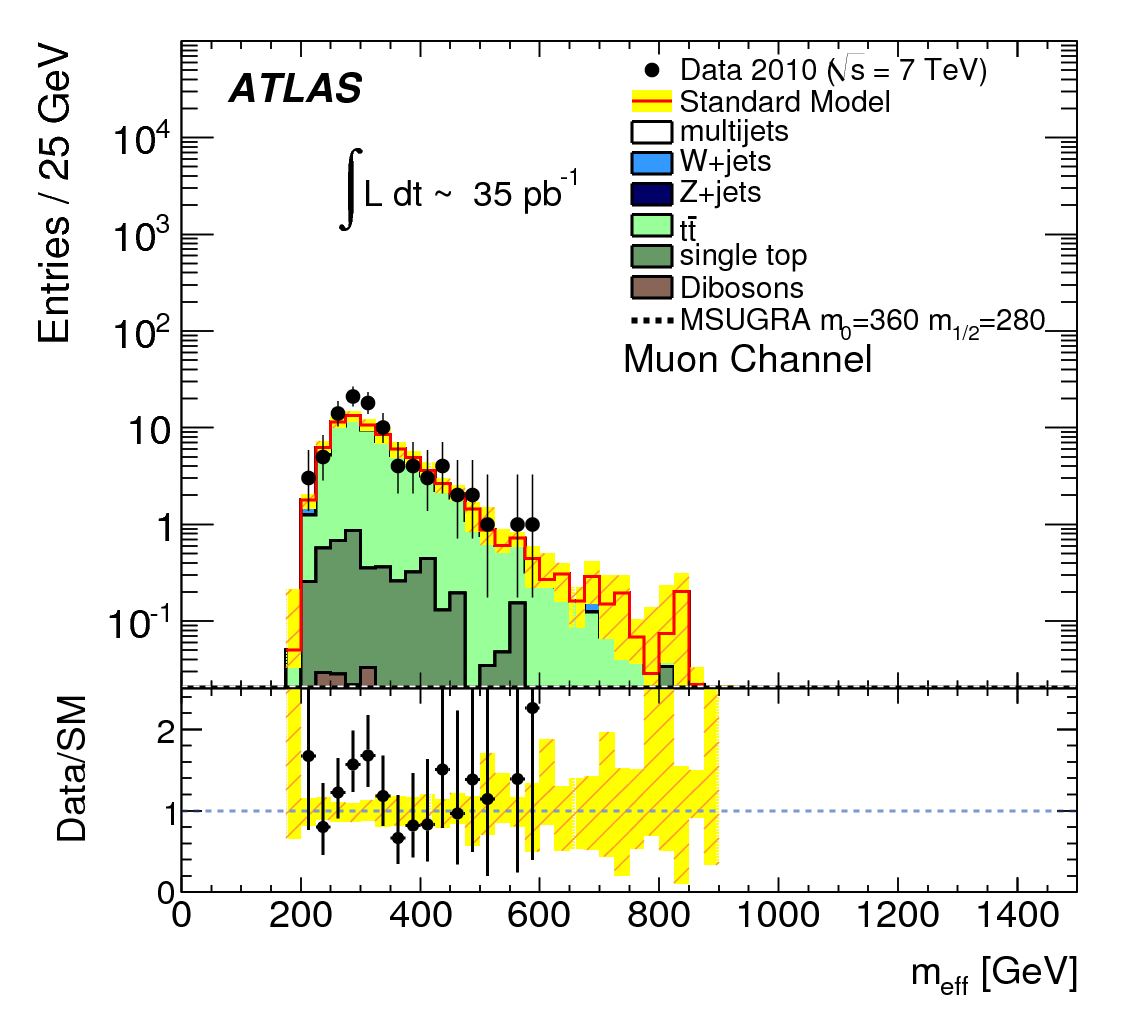} 
\end{center}
\caption{\label{fig:meff_distributions} The $M_{\rm{eff}}$ distributions in the \textit{W} region (left) and in the top region (right) in the muon channel. The yellow bands indicate the uncertainties on the MC predictions.}
\end{figure}

\noindent This method is validated by comparing data to MC in the relevant distributions for the extrapolation. Figure \ref{fig:meff_distributions} shows for example the $M_{\rm{eff}}$ distributions in the TR and in the WR in the muon channel. The good agreement between data and MC gives confidence in the method.

\section{QCD background}

\noindent In contrast to the W and top backgrounds, the QCD background is estimated by a matrix method. As mentioned above, the signal selection cuts (tight selection cuts) choose only events with exactly one isolated lepton. By relaxing the isolation requirement for the lepton a loose control sample enriched with QCD events can be defined. The events passing the tight selection cuts can be expressed as the sum of QCD events (``fake'') and non-QCD events (``real'') as in equation \ref{eq:QCDtight}. 

 \be
 \label{eq:QCDtight}
   N^{\rm{obs}}_{\rm{tight}} = N^{\rm{real}}_{\rm{tight}} + N^{\rm{fake}}_{\rm{tight}}
\ee

\noindent In the same way, the events passing the loose but not the tight selection cuts can be written as the sum of QCD and non-QCD events. By defining two efficiencies by 
$
\epsilon^{\rm{real/fake}} = \frac{ N^{ \rm{real/fake} }_{ \rm{tight} } }{ N^{ \rm{real/fake} }_{ \rm{loose} } } 
$
\noindent this sum can be written as:

  \be
 \label{eq:QCDloose}
   N^{\rm{obs}}_{\rm{loose~not~tight}} = (1/\epsilon^{\rm{real}} - 1)N^{\rm{real}}_{\rm{tight}} + (1/\epsilon^{\rm{fake}} - 1)N^{\rm{fake}}_{\rm{tight}} 
  \ee

\noindent Both equations \ref{eq:QCDtight} and \ref{eq:QCDloose} are solved for the QCD events passing the tight selection cuts in equation \ref{eq:QCDest}. The result gives the QCD events in the SUSY signal region.

\be
 \label{eq:QCDest}
  N^{\rm{fake}}_{\rm{tight}} = \frac{N^{\rm{obs}}_{\rm{loose~not~tight}} - (1/\epsilon^{\rm{real}} - 1)N^{\rm{obs}}_{\rm{tight}}}{1/\epsilon^{\rm{fake}} - 1/\epsilon^{\rm{real}}} 
\ee

\noindent This method was applied to the whole dataset of 2010 of $35~\rm{pb^{-1}}$ in the electron and in the muon channel, respectively. Thus, $\epsilon^{\rm{fake}} \sim 0.2 - 0.3$ was obtained in a QCD dominated control region with $E_{\rm{T}}^{\rm{miss}} < 40~\rm{GeV}$ and $M_{\rm{T}} < 40~\rm{GeV}$, whereas $\epsilon^{\rm{real}} \sim 0.9 - 1.0$ was taken from Monte Carlo. Only an upper limit on the QCD background in the signal region could be derived with $< 0.5$ events in the muon channel and $< 0.3$ events in the electron channel due to low statistics in the loose-but-not-tight events.

\section{Outlook}

\noindent The QCD, $t\bar{t}$ and \textit{W}+jets backgrounds in the signal region are estimated with the methods presented. Other backgrounds are taken from simulation. The final results are detailed in table \ref{table:results_1lepton}. The top background is with 1.76 $\pm$ 0.67 (muon channel) and 1.34 $\pm$ 0.52 (electron channel) the most dominant background in the signal region. In total, $2.25 \pm 0.94$ background events are expected in the muon channel and $1.81 \pm 0.75$ background events in the electron channel, but only 1 event passes all the signal selection cuts presented above in each of the electron and the muon channels. An interpretation of the results in terms of limits is given elsewhere \cite{1-lepton_paper}.

\begin{table}[t]
\caption{\label{table:results_1lepton} The number of observed events in $\int L dt = 35 ~\rm{pb^{-1}}$ is compared to the total number of background events expected. The contribution of $t\bar{t}$, W and Z events and QCD events to the total number of background events is given.}
\vspace{0.4cm}
\begin{center}
\resizebox{\textwidth}{!}{
\begin{tabular}{|c|c|c|ccc|}
\hline
channel & observed events & sum estimated BG events & estimated top & estimated WZ & estimated QCD\\\hline
   muon & 1 & 2.25 $\pm$ 0.94 & 1.76 $\pm$ 0.67 & 0.49 $\pm$ 0.36 & $0.0^{+0.5}_{-0.0}$\\
   electron & 1 & 1.81 $\pm$ 0.75 & 1.34 $\pm$ 0.52 & 0.47 $\pm$ 0.40 & $0.0^{+0.3}_{-0.0}$\\ 
\hline
\end{tabular}
}
\end{center}
\end{table}

\end{document}